%% file: main.tex
\documentclass[runningheads]{llncs}
\usepackage[english]{babel}
\usepackage[utf8]{inputenc}
\usepackage{amsmath}
\usepackage{graphicx}
\usepackage[colorinlistoftodos]{todonotes}
\usepackage{graphicx,multicol,latexsym,amsmath,amssymb}
\usepackage[]{subfig}
\usepackage[printwatermark]{xwatermark}
\begin{document}
\title{Impact of Indirect Contacts in Emerging Infectious Disease on Social Networks}
\titlerunning{Impact of indirect contacts in emerging disease}
%
\author{Md Shahzamal\inst{1,2} \and
Raja Jurdak\inst{2,1} \and Bernard Mans\inst{1}\and
Ahmad El Shoghri\inst{2,3} \and Frank De Hoog\inst{2}}
\authorrunning{M. Shahzamal et al.}
%
\institute{Macquarie University, Sydney, Australia\\
\email{\{md.shahzamal@students,bernard.mans@\}mq.edu.au}\and
Data61, CSIRO, Australia\\
\email{\{raja.jurdak,frank.dehoog\}@data61.csiro.au} \and
University of New South Wales, Sydney, Australia\\
\email{ahmad.elshoghri@students.unsw.edu.au}}
\maketitle             
\begin{abstract}
Interaction patterns among individuals play vital roles in spreading infectious diseases. Understanding these patterns and integrating their impact in modeling diffusion dynamics of infectious diseases are important for epidemiological studies. Current network-based diffusion models assume that diseases transmit through interactions where both infected and susceptible individuals are co-located at the same time. However, there are several infectious diseases that can transmit when a susceptible individual visits a location after an infected individual has left. Recently, we introduced a diffusion model called same place different time (SPDT) transmission to capture the indirect transmissions that happen when an infected individual leaves before a susceptible individual's arrival along with direct transmissions. In this paper, we demonstrate how these indirect transmission links significantly enhance the emergence of infectious diseases simulating airborne disease spreading on a synthetic social contact network. We denote individuals having indirect links but no direct links during their infectious periods as hidden spreaders. Our simulation shows that indirect links play similar roles of direct links and a single hidden spreader can cause large outbreak in the SPDT model which causes no infection in the current model based on direct link. Our work opens new direction in modeling infectious diseases.  

\end{abstract}

\input{Introduction}
\input{riskmodel}
\input{Networkmodel}
\input{simulation}
\input{discussion}
\bibliographystyle{unsrt}

\bibliography{reflist}
\end{document}

%% file: Introduction.tex
\section{Introduction}
Analysis of social contact networks is critical to understand and model diffusion processes within these networks. The contact patterns among individuals significantly impact spreading dynamics. Thus, a large body of work has attempted to reveal interactions among the network properties: e.g., temporal properties, burstiness and repetitive behaviors of contacts and the diffusion dynamics. Most of these works assume that the interacting individuals are co-located in the same physical or virtual space at the same time to create a link~\cite{holme2015modern,pastor2015epidemic}. However, this assumption does not hold in many diffusion processes such as airborne infectious disease spreading, vector borne disease spreading, and posted message diffusion in online social networks~\cite{fernstrom2013aerobiology,gruhl2004information}. In these diffusion processes, the recipient may receive the spreading items from the sender without concurrent presence if the spreading item survives in the deposited location after its generation. This process can be explained with an example of airborne disease spreading where an infected individual deposits infectious particles at the locations where they visit. These particles can transmit to susceptible individuals who visit the locations even after the infected individual leaves as the airborne infectious particles suspend in the air for a long time~\cite{fernstrom2013aerobiology,shahzamal2017airborne}. Therefore, susceptible individuals do not need to be in the same place at the same time with the infected individual to contract disease.

Our recent work introduced a diffusion model called same place different time (SPDT) to capture such diffusion processes, e.g., airborne disease spreading. In the SPDT model, the transmission link is created between two individuals for visiting the same location within a specified time window~\cite{shahzamal2017airborne}. For example, the infectious particles can transmit from the infected individual who arrived in a location to susceptible individuals who are present in that location or who arrive later on. Here, the created link is directional from the infected to the susceptible individual. We call these links SPDT links with components: (1) direct link when both individuals are present at the location; and/or (2) indirect link when the infected individual has left, but the susceptible individual is still present in the location or arrives later on. The transmission capability of SPDT links depends on the environmental conditions such as temperature and wind-flow etc. which determine the particle removal rate~\cite{fernstrom2013aerobiology,brenner2017climate,shahzamal2017airborne}. Thus, an SPDT link is characterized by the arrival and departure timings of infected and susceptible individuals and environmental conditions.

In the literature, previous works have aimed to characterize diffusion dynamics based on the interaction mechanisms among individuals~\cite{holme2015modern,zhang2018human,meyer2017incorporating}. These works have studied various aspects of human contact patterns from microscopic properties such as temporal behavior of contacts, burstiness, inter-event time and repetitiveness to the higher level structures such as clustering and community formation among individuals. The microscopic properties control the higher level structures of social contact networks and hence strongly influence the diffusion dynamics on it~\cite{shirley2005impacts,min2011spreading}. Thus, inclusion of indirect links in the SPDT model may modify the higher level structures that are present in the current same place same time (SPST) based individual to individual contact networks and influence the diffusion dynamic significantly. To the best of our knowledge, this work is the first to investigate the impact of indirect links occurring at the individual level through simulation of airborne disease spreading on social contact network.

To study the impact of indirect links, it is required to collect the sufficiently dense individual level interaction data with high spatial and temporal resolutions. However, it is quite difficult to gather such data for a population over a sufficient period of time due to privacy issues and the complexity of collection methods. Thus, we use synthetic traces generated by our SPDT network model that provides the SPDT links among the nodes present in the network. This model is fitted with the real SPDT contact network properties found among the users of social networking application Momo~\cite{chen2013and}. The generated SPDT links contain the timing information of nodes' interactions mimicking the arrival and departure of individuals in a location. Thus, SPDT links easily allow quantification of infection risk with environmental factors if neighbor nodes of a link are infected. The infection risk for concurrent interaction between susceptible and infected individuals is formulated in~\cite{issarow2015modelling}. We improve this model to find infection risk for SPDT interaction. In our simulations, disease propagates on networks following the Susceptible-Exposed-Infected-Recovered (SEIR) epidemic model.

Our main goal is to determine the importance of indirect links for spreading infections. We first study diffusion dynamics on both the SPDT and the SPST networks selecting the seed nodes (initiating the spreading process) according to SPDT link components during their infection periods: having only indirect links, hidden spreaders, or both direct and indirect links. We gradually increase the proportion of hidden spreaders and find their impact on the diffusion, and highlight the inability of current models in capturing the effect of hidden spreaders on diffusion. We next explore how the changes in network properties that arise from the inclusion of indirect links impact the diffusion. We consider nodes creating small number of direct links when they are infected as seed nodes and look at the emergence of disease from each single seed node in  both the SPST and the SPDT networks. Finally, we study the potential for disease emerge by a hidden spreader acting as a single seed node.

The rest of the paper is organized as follows. In Section 2, we present the improved infection risk assessment model for SPDT links. Section 3 describes our methodology. The experimental setup and analysis of results are presented in Section 4. Section 5 concludes our work and provides future research directions.

%% file: riskmodel.tex
\section{Infection Risk in SPDT}
In this section, we present the methods of determining the infection risk for a susceptible individual that has SPDT links with the infected individuals. When an infected individual appears at a location $L$, he deposits infectious particles in the proximity of $L$. The number of infection particles $n$ deposited (through coughing) per second by an infected individual into the proximity is
\begin{equation}
n=0.2 fvc
\end{equation}
where $f$ is the coughing frequency (coughs/second), $v$ is the volume of each cough (m$^{3}$) and $c$ is the concentration of infectious particles in the cough droplets (particles/m$^{3}$). If the particles are removed from the space of volume $V$ with a rate $r$, the accumulation rate of particles in the proximity can be given by 
\begin{equation*}
V\frac{\mathrm{d}N }{\mathrm{d} t}=n-Nr
\end{equation*}
where $N$ is the current number of particles at $L$ and $r=(1-b)(1-g)$, b is the infectivity decay rate of particles and g is the air exchange rate from $L$~\cite{issarow2015modelling}. The particle concentration at time $t$ after the infected individual arrives at $L$ is given as 
\begin{equation*}
\int_{0}^{N_t}\frac{dN}{n-Nr}=\frac{1}{V}\int_{0}^{t}dt
\end{equation*}
This leads to
\begin{equation}
N_t=\frac{n}{r}\left(1-e^{-\frac{rt}{V}}\right)
\end{equation}
If a susceptible individual stays within the proximity of $L$ from $t_c$ to $t_c+t_d$ for a period $t_d$ when the infecter is concurrently present at $L$, the number of particles inhaled by the susceptible individual with pulmonary rate $q$ for this direct link is
\begin{equation*}
E_d=\frac{qn}{r}\int_{t_c}^{t_c+t_d}\left(1-e^{-\frac{rt}{V}}\right) dt
\end{equation*}

\begin{equation}
=\frac{qn}{r}\left[\left(t_d+t_c + \frac{V}{r}e^{-\frac{r(t_c+t_d)}{V}}\right)-\left(t_c+\frac{V}{r}e^{-\frac{rt_c}{V}}\right)\right]
\end{equation}

If the susceptible individual stays with the infected individual as well as after the latter leaves $L$, it will have both direct and indirect transmission links. The number of particles inhaled by the susceptible individual due to the direct link within the time $t_c$ and $t_a$ is given by
\begin{equation}
E_d=\frac{qn}{r}\left[\left(t_a+\frac{V}{r}e^{-\frac{rt_a}{V}}\right)-\left(t_c+\frac{V}{r}e^{-\frac{rt_c}{V}}\right)\right]
\end{equation}
where $t_a$ is the stay duration of infected individual. For the indirect link from time $t_a$ to $t_c+t_d$, we need to compute the particle concentration during this period which decreases after the infected individual leaves. The particle concentration at time $t_a$ can be given as
\begin{equation*}
N_a=\frac{n}{r}\left(1-e^{-\frac{rt_a}{V}}\right)
\end{equation*}
The particle concentration at time $t$ after the susceptible leaves the proximity at time $t_a$ is given by
\begin{equation*}
V\frac{dN}{dt}=-Nr
\end{equation*}
Thus, the concentration at time $t$ will be
\begin{equation*}
\int_{N_a}^{N_t}\frac{dN}{N}=-\frac{r}{V}\int_{t_a}^{t} dt
\end{equation*}
Thus, we have
\begin{equation*}
N_t=N_a e^{-\frac{r}{V}(t-t_a)}=\frac{n}{r}\left(1-e^{-\frac{rt_a}{V}}\right)e^{-\frac{r}{V}(t-t_a)}
\end{equation*}
The susceptible individual inhales particles during the indirect period from $t_a$ to $t_c+t_d$, quantified by
\begin{equation}
E_i=\int_{t_a}^{t_c+t_d}qN_t dt =-\frac{nqV}{r^2}\left(1-e^{-\frac{rt_a}{V}}\right)\left[e^{-\frac{r}{V}(t_c+t_d-t_a)}-1\right]
\end{equation}
If the susceptible individual is only present for the indirect period at the proximity, the number of inhaled particles for the period from $t_c$ to $t_c+t_d$ is given
\begin{equation}
E_i=\int_{t_c}^{t_c+t_d}qN_t dt =-\frac{nqV}{r^2}\left(1-e^{-\frac{rt_a}{V}}\right)\left[e^{-\frac{r}{V}(t_c+t_d-t_a)}-e^{-\frac{r}{V}(t_c-t_a)}\right]
\end{equation}
Thus, the total inhaled particles can be given for a susceptible individual by
\begin{equation}\label{eq:expo}
E=E_d + E_i
\end{equation}
The equations determine the received exposure for one SPDT link with an infected individual, comprising both direct and indirect links. If a susceptible individual has $m$ SPDT links during an observation period $T$, the total exposure is 
\begin{equation*}
E_{T}=\sum_{k=0}^{m}E_{k}
\end{equation*}
where $E_k$ is the received exposure for k$^{th}$ link. The probability of infection for causing disease can be determined by the dose-response relationship defined as 
\begin{equation}\label{eq:prob}
P(I=1)=1-e^{-\sigma E_T}
\end{equation}
where $\sigma$ is the probability that an infectious particle reaches to the respiratory tract and initiates infection~\cite{jones2015dose}. It is assumed that inhaling one infectious particle has $50\%$ chance of contracting the disease~\cite{teunis2010high}. Therefore, we can calculate $\sigma$ as
\[0.5=1-e^{-\sigma}\]
\[\sigma=0.693\]
In this risk formulation, $\sigma$ is homogeneous for all susceptible individuals.

%% file: Networkmodel.tex
\section{Methodology}
For studying the impact of indirect contacts on the spreading process, individual level interaction data is required. However, it is difficult to collect such data with sufficiently high contact density, and with high spatial and temporal resolution. Thus, we generate synthetic individual to individual interaction data using our developed SPDT network model. This network model includes indirect links of disease transmission along with direct links in creating SPDT link between two nodes. Then, the SEIR epidemic model is simulated on this network with disease parameters of airborne diseases. In this section, we describe our methodology. 

\subsection{Contact Network}
A network of M nodes is constructed following the approach of activity-driven temporal network generation where nodes switch between active periods and inactive periods over the simulation period of $T$ discrete time steps. Active periods $t_a$ mimic staying at a location and allow nodes to create SPDT links with other nodes while inactive periods $t_w$ represent time windows when a node does not create links to other nodes, but receives SPDT links. The SPDT link is directed from the host node (as infected) to the neighbor node (susceptible node). The duration of active periods $t_a=\{1,2,3,\ldots\}$ are randomly drawn from $P_1(t_a)\sim geometric(\lambda)$ with the scaling parameter $\lambda$. The inactive period durations $t_w=\{1,2,3,\ldots\}$ are also drawn from $P_2(t_w)\sim geometric(\rho)$ with the scaling parameter $\rho$. The value of $\lambda$ is constant for all nodes but $\rho$ is assigned heterogeneously to model different frequencies of individuals in visiting locations. The heterogeneous activation potentiality $\rho$ is drawn from a power law $F_1(\rho)\sim \rho^{-\alpha}$ with the scaling parameter $\alpha$.

Corresponding to each active period, an indirect transmission period $\delta$ is added with $t_a$ to capture the indirect links of disease transmission. During $t_a$ and corresponding $\delta$, a node creates a specific number of SPDT links $d$ given by
\begin{equation*} \label{actdgr}
Pr\left (d\right)=(1-\mu) \mu ^{d-1}
\end{equation*}
where $\mu$ is the propensity to access public places that is also drawn from the power law $F_2(\mu)\sim \mu^{-\beta}$ with the scaling parameter $\beta$. The variations in $d$ capture spatio-temporal dynamics of social networks. 

Each of these SPDT links has the timing characteristics: $t_a$ representing the time duration node stays at the interacted location, $t_c$ is the delay time of neighbor node arrival after the host node appeared, and $t_d$ is the time duration neighbor node (end nodes of a link) stays at the interacted location. The link creation delay $t_c$ is drawn from a truncated geometry distribution as follows as 
\begin{equation*} \label{ldelay}
Pr\left (t_{c}\right )=\frac{p_{c} \left(1-p_{c}\right)^{t_c}}{1 -(1-p_{c})^{t_{a}+\delta}} 
\end{equation*}
where $p_c$ is the probability of creating a link with a neighbor. The value of $t_d$ is drawn as $P_3(t_d)\sim geometric(p_b)$ with link breaking probability $p_b$. The value of $p_c$ and $p_b$ are constant for all nodes in the network.

Nodes maintain their social structure by applying their memory of previous contacts in selecting new neighbors with the probability $Pr(n_t +1)= \mu_i \theta/(n_t + \mu_i \theta)$, where $n_t$ is the number of nodes node $i$ has contacted up to time t. With greater public accessibility $\mu$, nodes will have higher contact set sizes. They also need to be selected as a neighbor by more nodes. Thus, the probability of being selected as a neighbor is $p(\mu_{j})=\mu_{j}/M\left\langle \mu \right\rangle$. The neighbor selection mechanism also considers that if some nodes have included node $j$ as neighbor, there is $\varphi$ chance to select them as neighbors by node $j$. 

The network model parameters are fitted with the real SPDT network constructed using the 2 million locations updates of 126K users collected over a week from the social networking application Momo. The location updates from Beijing city are applied to estimate the model parameters while the updates from Shanghai are used to validate the model. The capability of the model to simulate the SPDT diffusion process is also verified in detail. We generate synthetic SPDT networks to conduct our experiments using this model.

\subsection{Epidemic Model}
For propagating disease on the generated SPDT contact network, we  consider a compartment-based Susceptible-Exposed-Infected-Recovered (SEIR) epidemic model. In this model, nodes remain in one of the four compartments, namely, Susceptible (S), Exposed (E), Infectious (I) and Recovered (R). If nodes in the susceptible compartment receive SPDT links from the nodes in the infectious compartment, the former will receive the infectious pathogens for both direct and indirect periods and may contract the disease. At the beginning of contraction, a susceptible node moves to the exposed (E) state where it cannot infect others. The exposed node becomes infectious (I) after a latent period. The infectious node continues to infect other nodes connecting through the SPDT links over its infection period until they enter the recovered state R~\cite{stehle2011simulation}. It has been shown that the latent period is in the range over 1 to 2 days. Our simulations assume that if a node is infected in the current day of simulation, it starts infecting others on the next day of simulation. The infection period is shown to be in the range over 3 to 5 days for influenza-like diseases~\cite{chowell2011characterizing}. As the values can vary for each individual even for the same disease, we derive the parameters from a random uniform distribution within the observed empirical ranges.

%% file: simulation.tex
\section{Simulation and Analysis}
We conduct various simulation experiments to understand the impact of indirect transmissions in shaping the epidemic dynamics on the social contact network. We consider a network of 300K nodes generated by the SPDT network model for a 32 day period. In this network, nodes create SPDT links that are inclusive of direct and indirect transmission links. Removing the indirect links from the above network results in an SPST network with only direct links. Therefore, the SPDT and SPST networks include the same nodes, but their connectivity differs due to the presence or absence of indirect links. If a node is infected and creates only indirect links during its infection period, we refer to this node as a `{\em hidden spreader}'. The contribution of hidden spreaders to infections in the original SPST network is nil while they significantly contribute to spread in the SPDT network (and thus possibly promote new direct links). In this section, we explore the extent to which indirect links can impact the disease spread dynamics on contact networks through intensive simulations.

\subsection{Simulation Setup}
The generated synthetic SPDT links of 300K nodes provide the traces for running data-driven disease simulations over a period of $T=32$ days. The disease on this network propagates according to the SEIR epidemic model. The simulation at $T=0$ starts with some seed nodes that are randomly selected according to the requirements of experiments described below. The nodes' disease status are updated at the end of each simulation day. During each day of disease simulation, the received SPDT links for nodes are analyzed to find which nodes have received SPDT links from the infected neighbor nodes. Then, we calculate the particles inhaled by the susceptible neighbor node for each of these SPDT links according to Equation~\ref{eq:expo} and sum them up to find the infection probability by Equation~\ref{eq:prob}. I order to keep the simulations simple, we assume that the inhaled infection particles can infect susceptible individual with the same rate over the day. The volume $V$ of the proximity is fixed to a constant value assuming that the distance, within which a susceptible individual can inhale the infectious particles from an infected individual, is 40m~\cite{fernstrom2013aerobiology} and the particles will be available up to the ceiling height $h$ of $3$m. We assign the other parameters as follows: cough frequency $f=18/hour$, total volume of the cough droplets $v=6.7\times 10^{-3}$ $ml$, pathogen concentration in the respiratory fluid $c=3.7\times 10^6$ $pathogens/m^3$, and pulmonary rate $q=7.5 $ $liter/min$~\cite{loudon1967cough,yin2012retrospective}. For each link, the infectivity decay rate $b$ and air exchange rate $g$ are selected randomly to find particles removal rate $r=(1-b)(1-g)$. The value of $b$ is randomly drawn from the range $(0.005,0.05) min^{-1}$ with a specified mean according to the experiments while $g$ is randomly drawn from the range $(0.25,5)h^{-1}$ with a specified median as the experiments require. The daily simulation outcomes are obtained for the epidemic parameters: the number of new infections, the disease prevalence as the number of current infections in the system and the cumulative number of infections. The statistics of these parameters provide the results of our experiments.

\subsection{Results and Analysis}
In the simulation of disease on the SPDT network, the infected nodes can be divided into two groups based on the SPDT links they form during their infection periods: 1) nodes with both direct and indirect links; and 2) nodes with only indirect links, i.e. hidden spreaders. A hidden spreader has zero infection force in the SPST network but can cause disease in the SPDT network. Thus, the increase in the proportion $P$ of hidden spreaders in the seed nodes set will reduce spreading speed in the SPST network while spreading speed is sustained for SPDT according to the potentiality of the indirect links. In our first experiment, the simulations begin with 200 seed nodes where we randomly pick 200P seed nodes from the hidden spreader set and (1-P)200 seed nodes from the non-hidden spreader set. We vary the value of P from 0 to 1 with the step 0.1. The seed nodes start infecting at $T=0$ and continue infecting for the period of days picked up randomly from the range (1, 5). With this set of seed nodes, we run simulations on the both SPST and SPDT networks and repeat 200 times for each value of P. We set the mean value of b to $0.01min^{-1}$ and the median of g to 1$h^{-1}$.
\begin{figure}
	\vspace{-0.6 em}
    \centering
	\subfloat{\includegraphics[width=0.24\textwidth, height=2.8 cm]{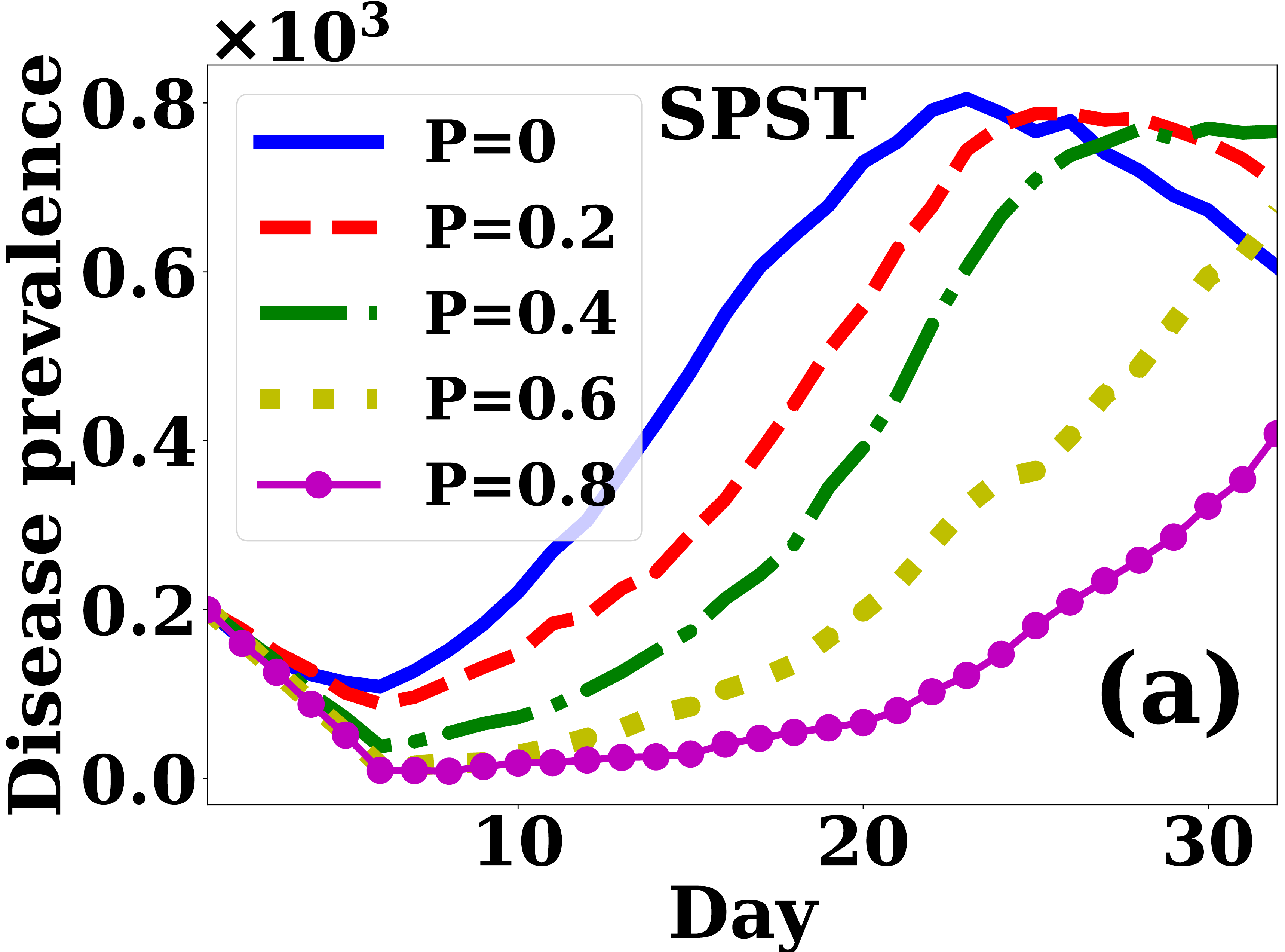}}~
    \subfloat{\includegraphics[width=0.24\textwidth, height=2.8 cm]{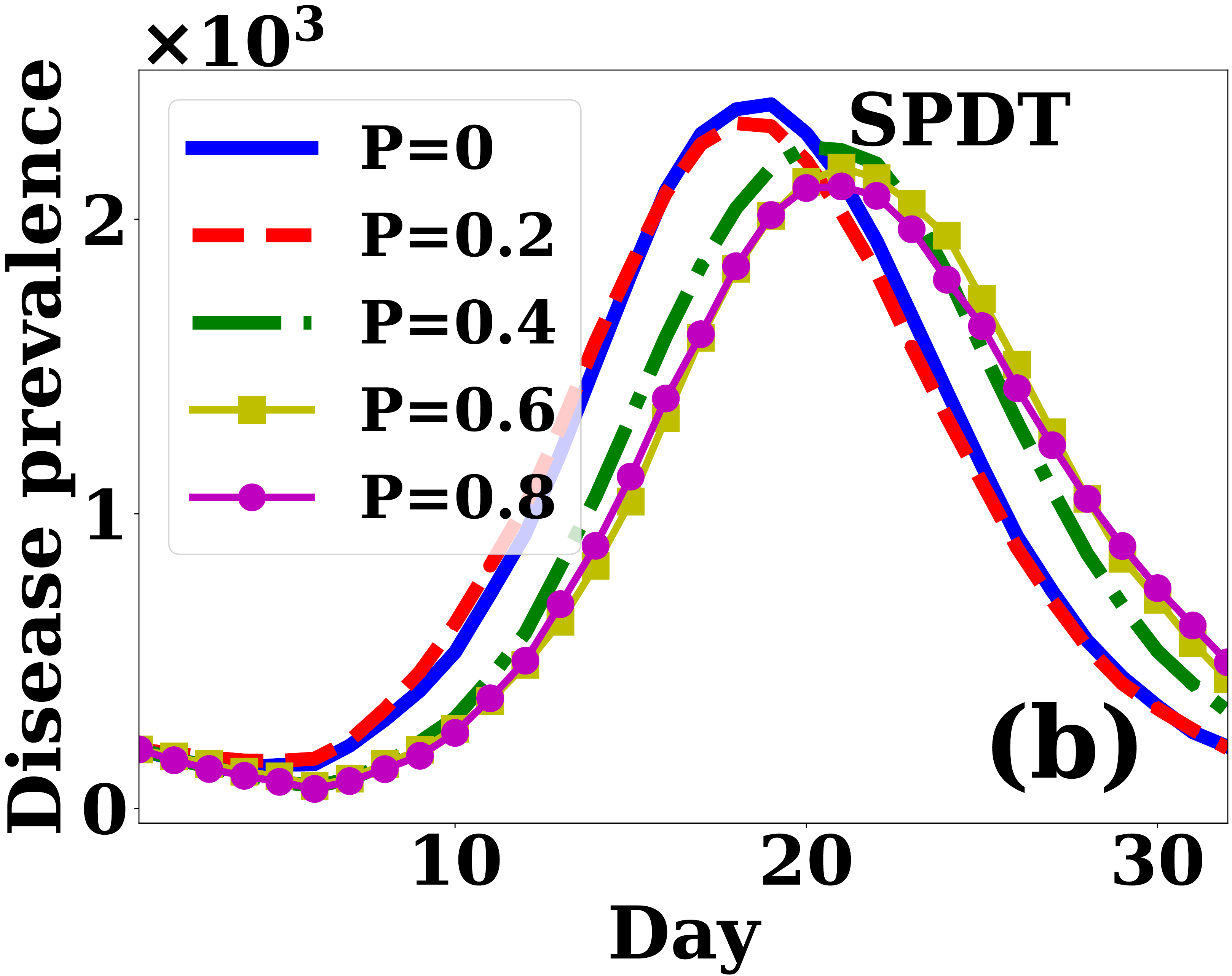}}~     	 \subfloat{\includegraphics[width=0.24\textwidth, height=2.8 cm]{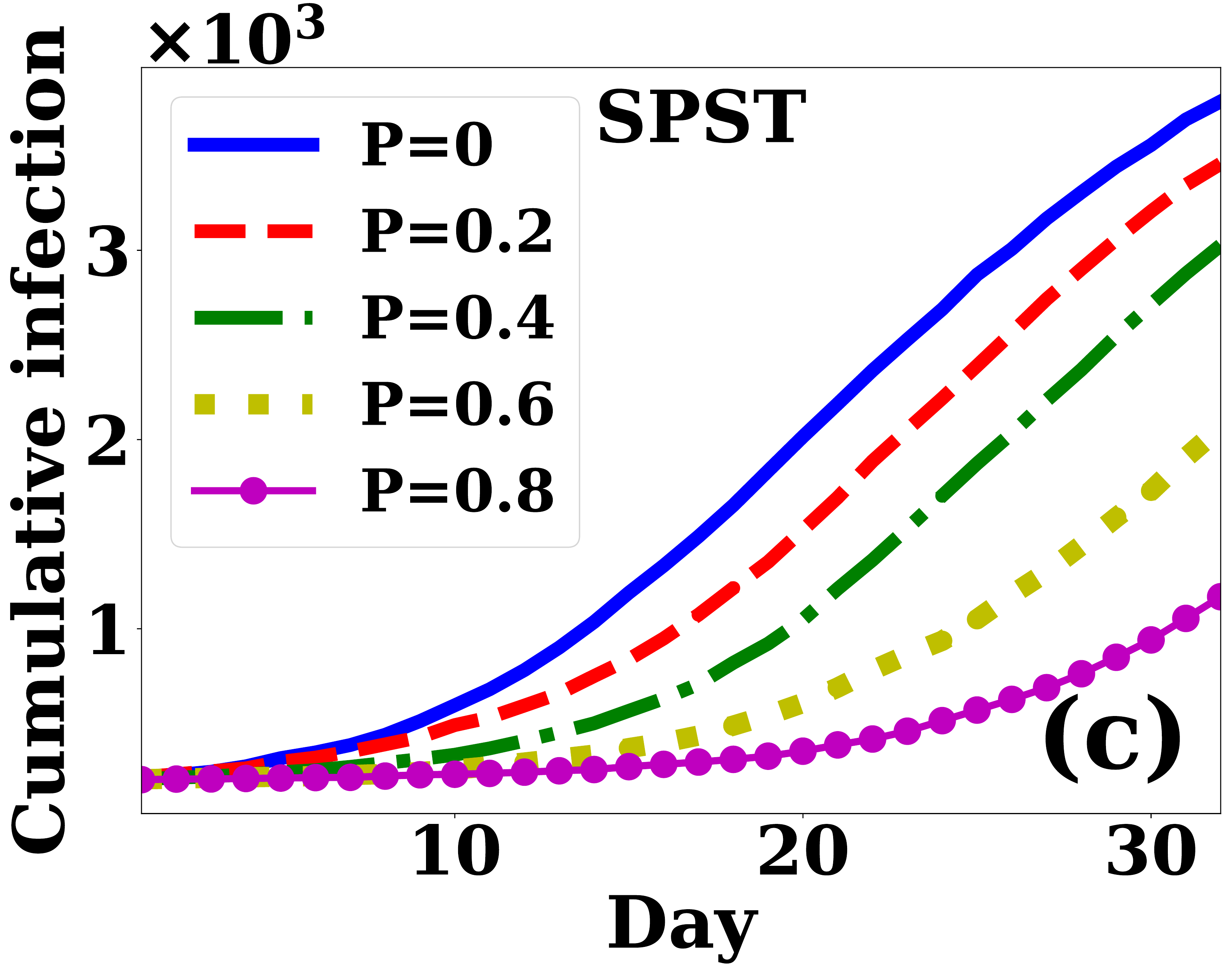}}~
     \subfloat{\includegraphics[width=0.24\textwidth, height=2.8 cm]{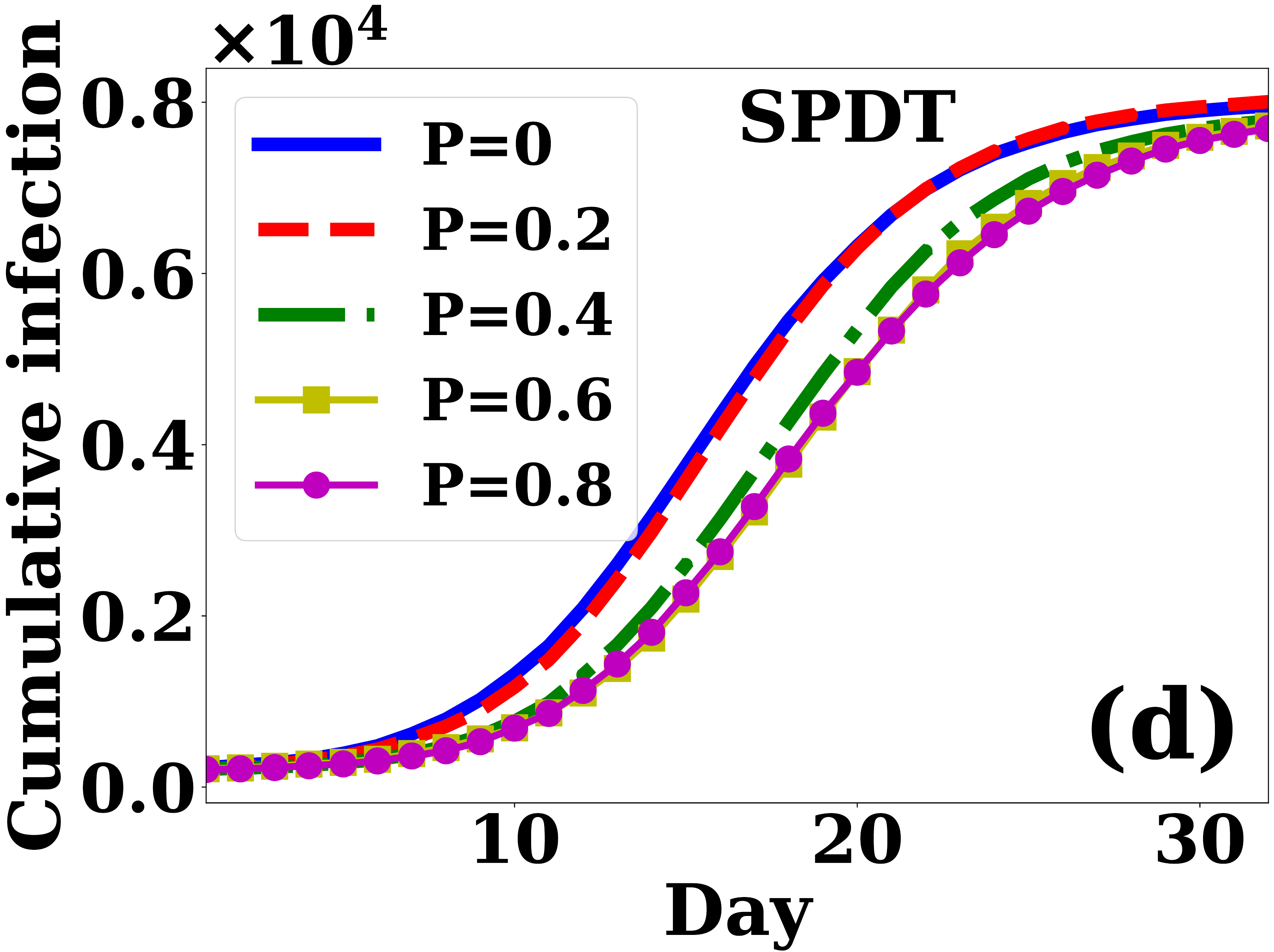}}
     \vspace{-0.6 em}
     \caption{Disease diffusion dynamics based on the disease prevalence~(a, b) and cumulative infection~(c, d) for different P in  both SPST and SPDT networks} 
     \label{fig:expa}
     \vspace{-1.4 em}
\end{figure}
\begin{figure}
    \vspace{-0.5 em}
	\centering
	\subfloat{\includegraphics[width=0.24\textwidth, height=2.75 cm]{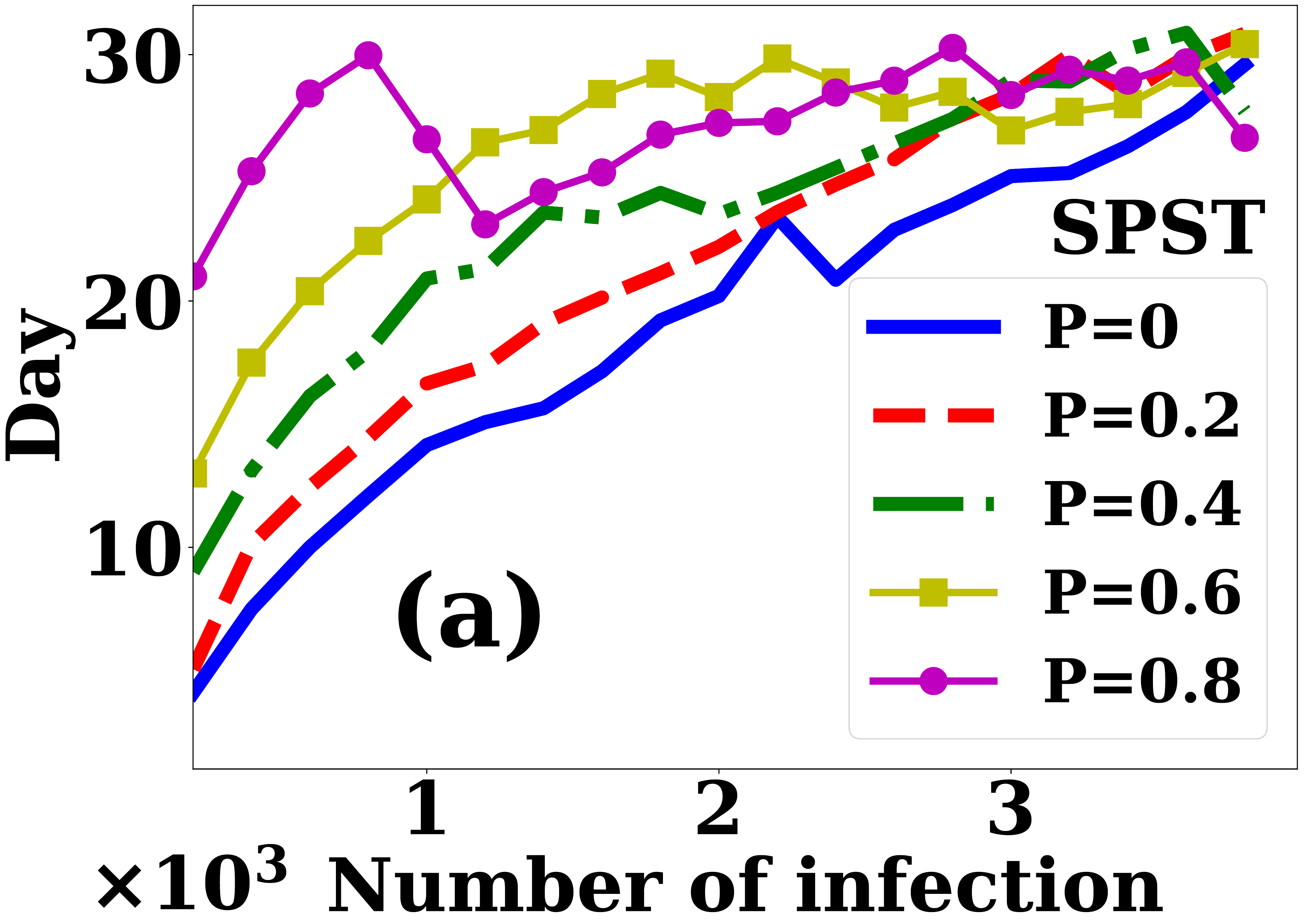}}~
    \subfloat{\includegraphics[width=0.24\textwidth, height=2.75 cm]{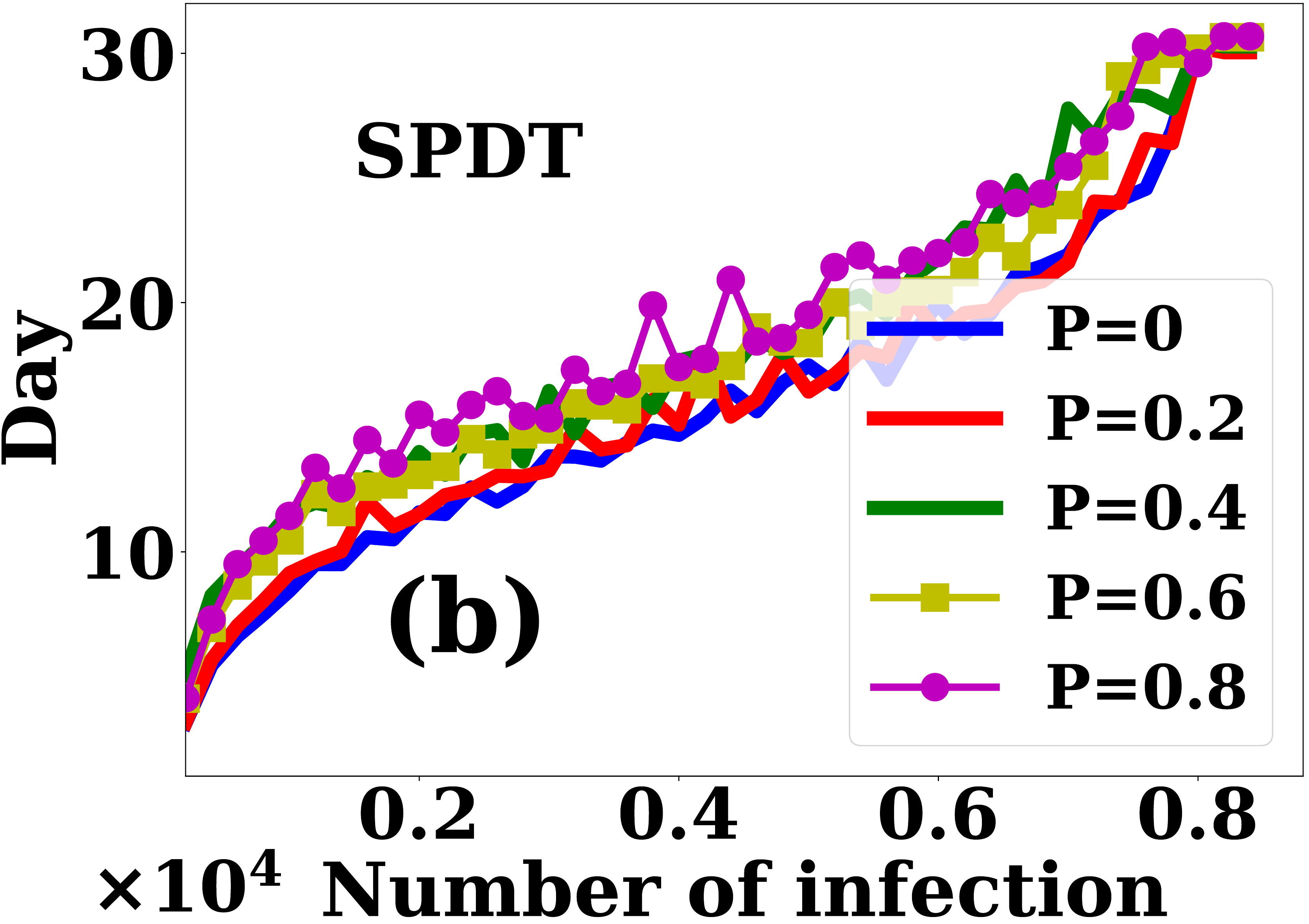}}~
     \subfloat{\includegraphics[width=0.25\textwidth, height=2.95 cm]{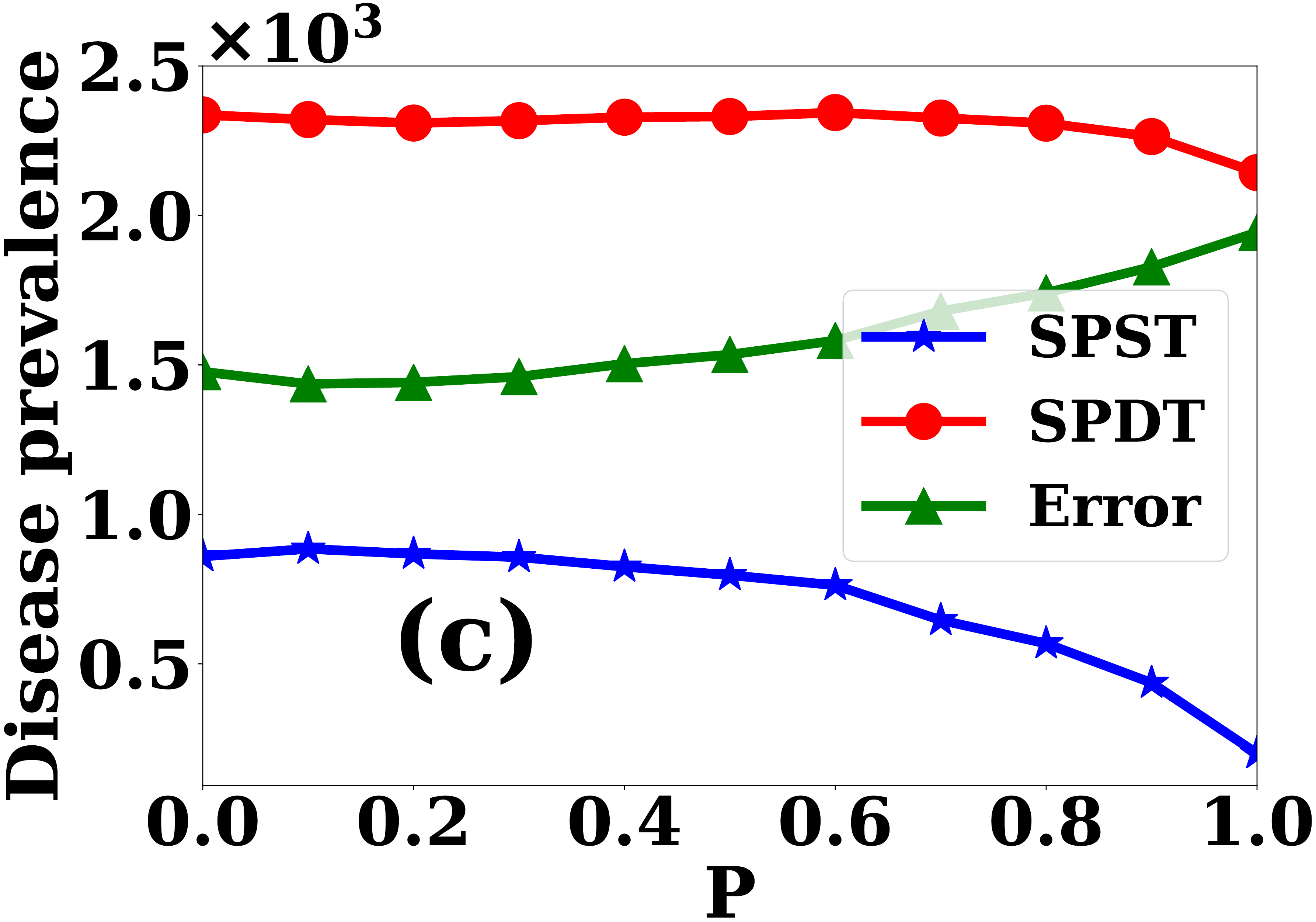}}~
     \subfloat{\includegraphics[width=0.25\textwidth, height=2.95 cm]{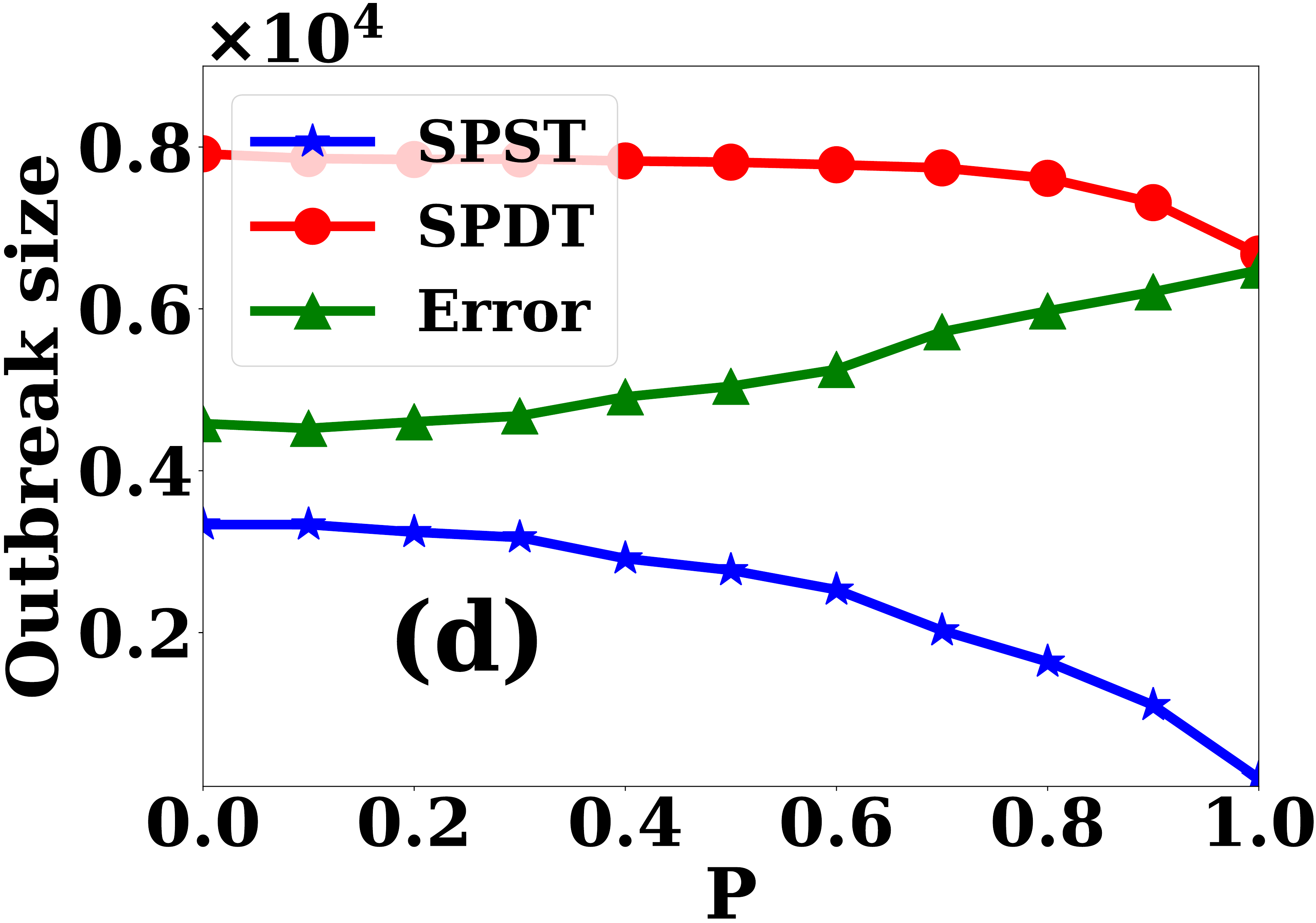}}
     \vspace{-0.2 em}
     \caption{Disease prediction performances for changes in P: (a, b) days requires to cause a specific number of infection since simulation starts and (c, d) error as differences between same metrics in the SPST and SPDT networks} 
     \label{fig:expb}
     \vspace{-1.9 em}
\end{figure}

Figure~\ref{fig:expa} shows the changes in the disease spreading dynamics, averaged over the 200 simultation runs, for P=$\{0,0.2,0.4,0.6,0.8\}$. For the SPST network, both prevalence and cumulative infections shrink with increasing P~(see Fig.~\ref{fig:expa}a and Fig.~\ref{fig:expa}c), as the increased proportion of hidden spreaders as seed nodes reduces the likelihood of seed nodes to trigger spreading. We also observe that the rate of reduction in both prevalence and cumulative infections increases with P, where the shrinking set of non-hidden spreaders among seed nodes reduces the likelihood of spread more rapidly. The disease prevalences and outbreak sizes decrease significantly for increasing P and dropping rates increase at the higher P's. On the other hand, the disease spreading behaviors do not change much with P in the SPDT network (see Fig.~\ref{fig:expa}b and Fig.~\ref{fig:expa}d). There is a slight time shift in prevalence with increasing P, yet the size of the epidemic remains similar. Noting that changes in P result in different proportions of hidden spreaders, the SPDT results confirm that the potency of hidden spreaders is almost similar to non-hidden spreaders in determining diffusion outcomes. 

We now explore disease prediction performance with changing P in  Figure~\ref{fig:expb} for delving deeper into the contribution of hidden spreaders. Figures~\ref{fig:expb}(a) and (b)  show the number of days required for causing a specific number of infections and how it is delayed with changing P. The SPST network fails to predict infection dynamics causing more delays to reach a specific number of cumulative infections as P increases (see Figure~\ref{fig:expb}(a)). The cumulative infections reaches 1000 by day 10 at P=0 but it is delayed to day 30 when P is 0.8. However, the required days to reach a number of cumulative infections changes slightly for the SPDT network in Figure~\ref{fig:expb}b with changing P as the indirect links have similar impact as direct links. The differences in outbreak sizes and disease prevalences between SPST and SPDT networks with the changes in P are shown in Figures~\ref{fig:expb}(c) and (d). As P changes from 0 to 1, the outbreak sizes in SPST network drop by 100\% from 3745 to 0 infections. By comparison, the outbreak size drop by 2.7\% changes from 7834 to 7625 infections in the SPDT network. Thus, the differences in outbreak sizes between SPST and SPDT networks vary from 4000 infections for P=0 to 7625 for P=1. This indicates that the underestimation of disease dynamics by SPST model increases with P, i.e. if the disease starts with  all hidden spreaders, it shows no emergence of disease.
\begin{figure}
\vspace{-1.7 em}
	\subfloat{\includegraphics[width=0.245\textwidth, height=2.7 cm]{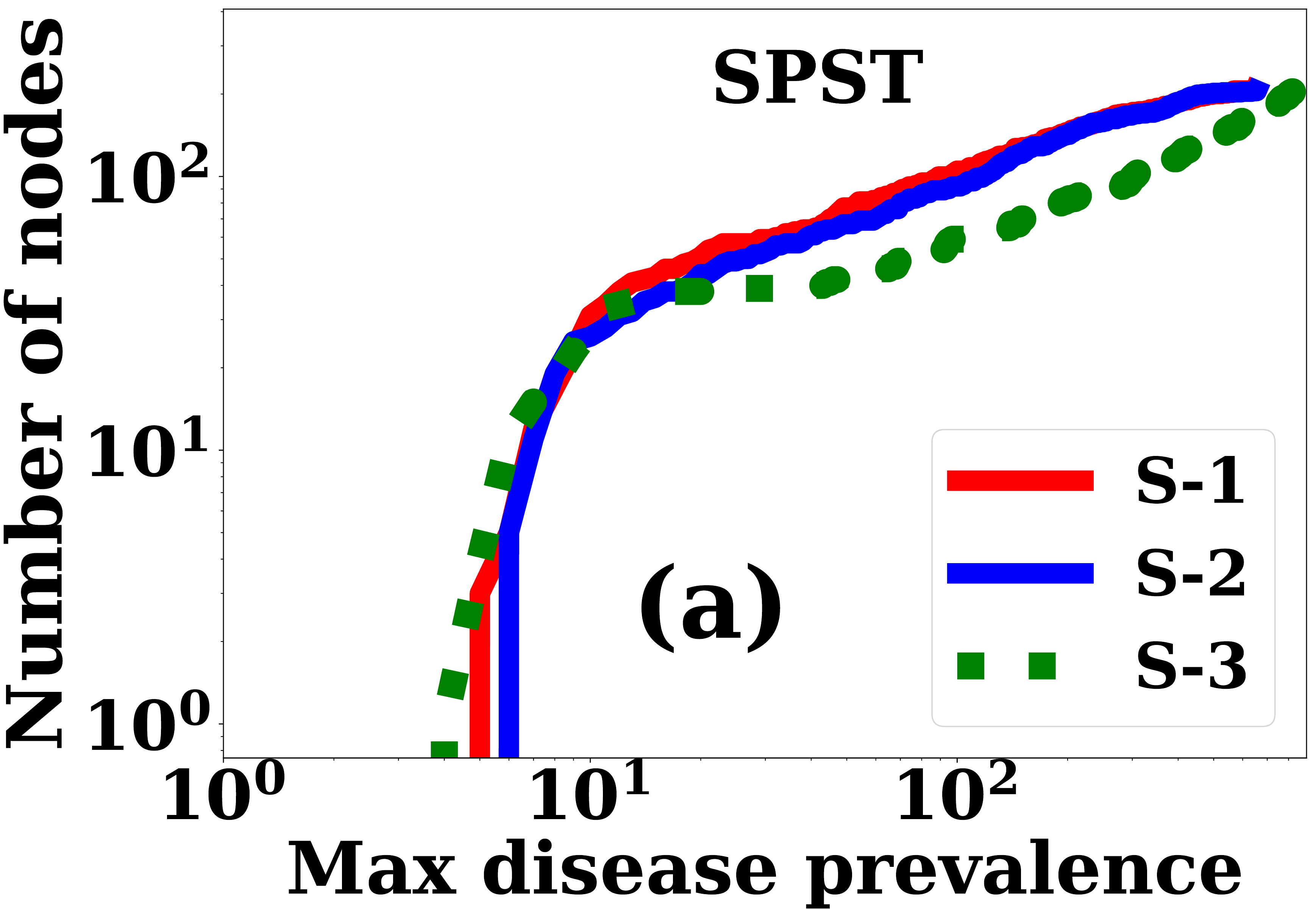}}~
    \subfloat{\includegraphics[width=0.245\textwidth, height=2.7 cm]{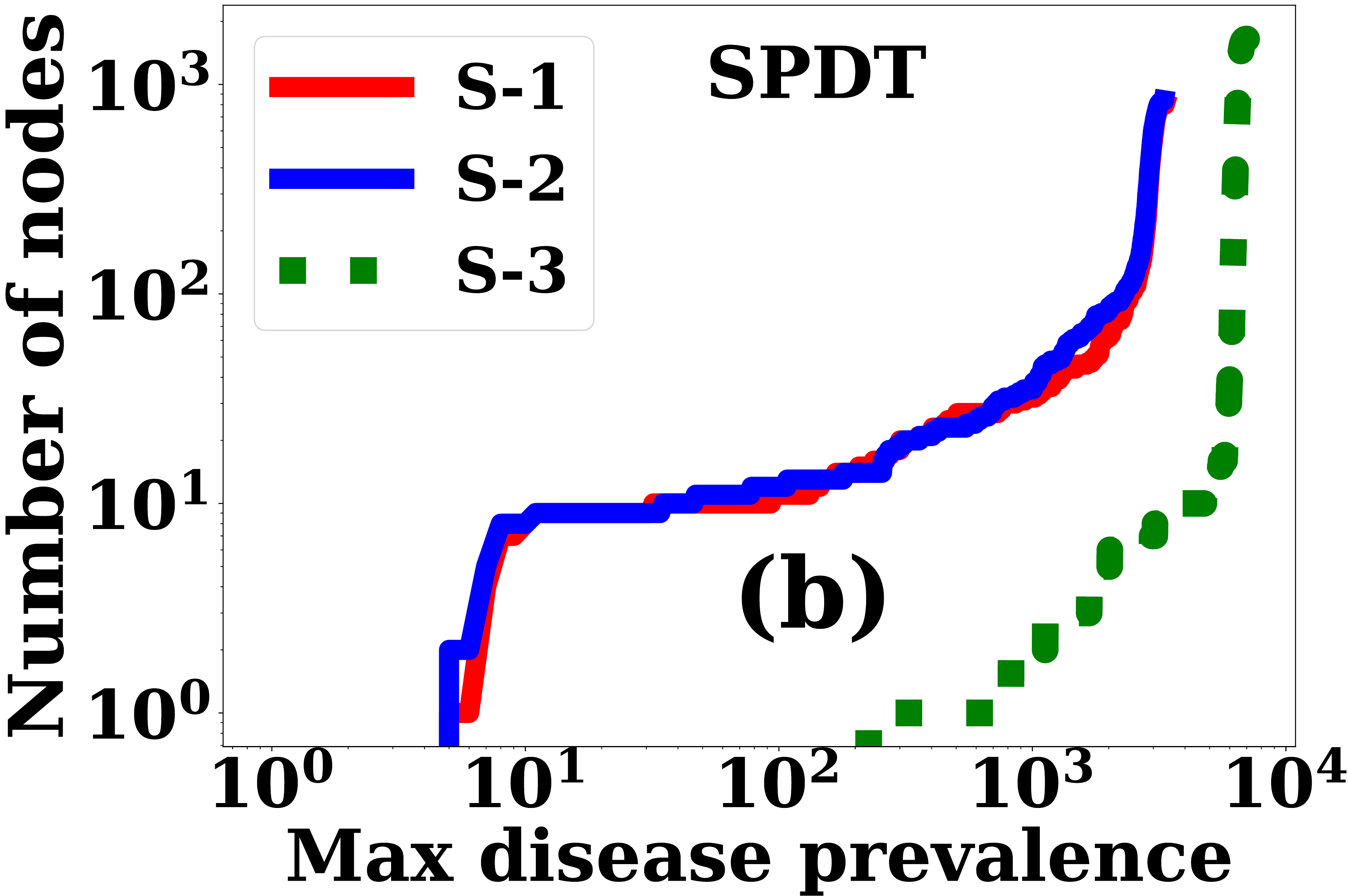}}~
     \subfloat{\includegraphics[width=0.245\textwidth, height=2.7 cm]{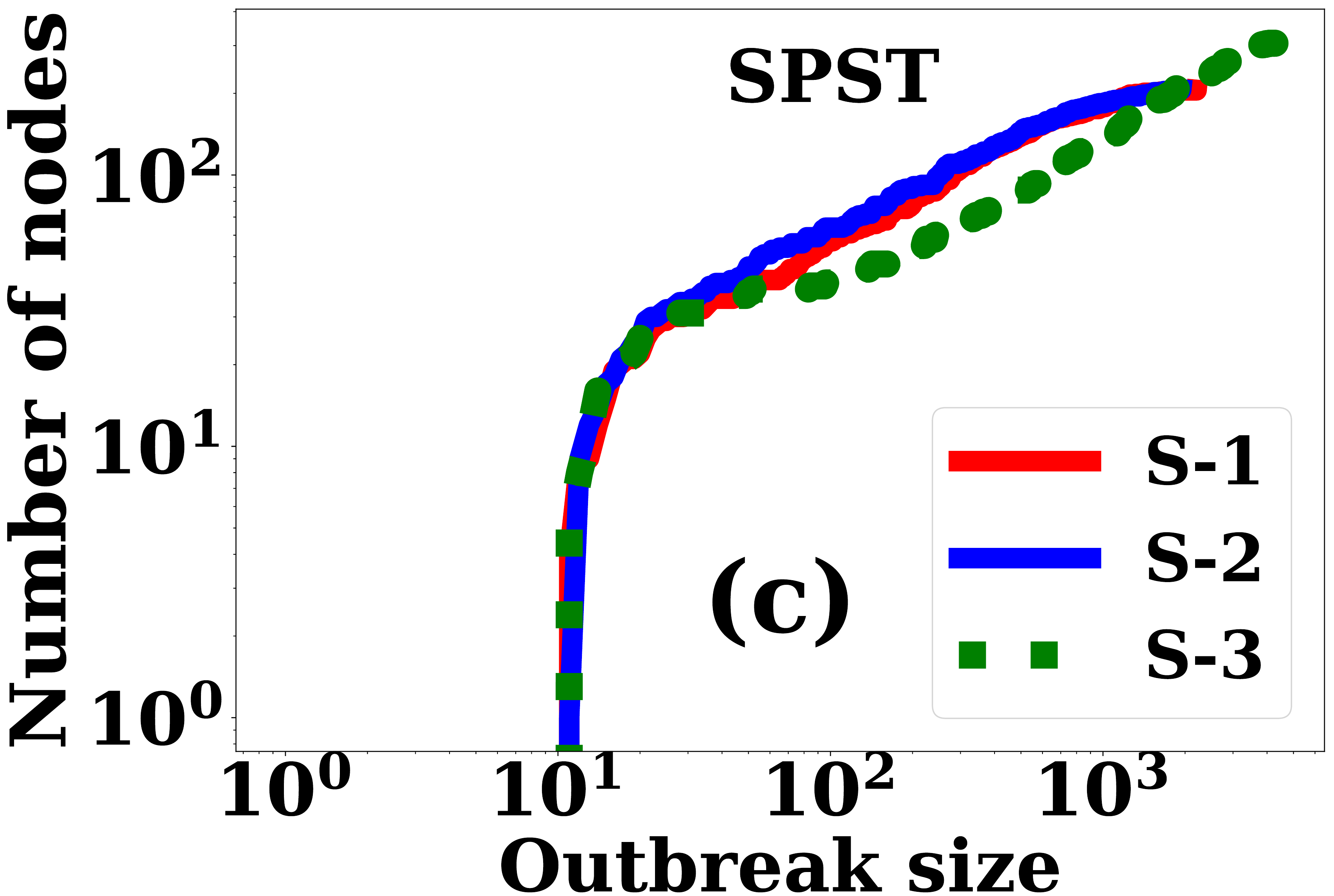}}~
     \subfloat{\includegraphics[width=0.245\textwidth, height=2.7cm]{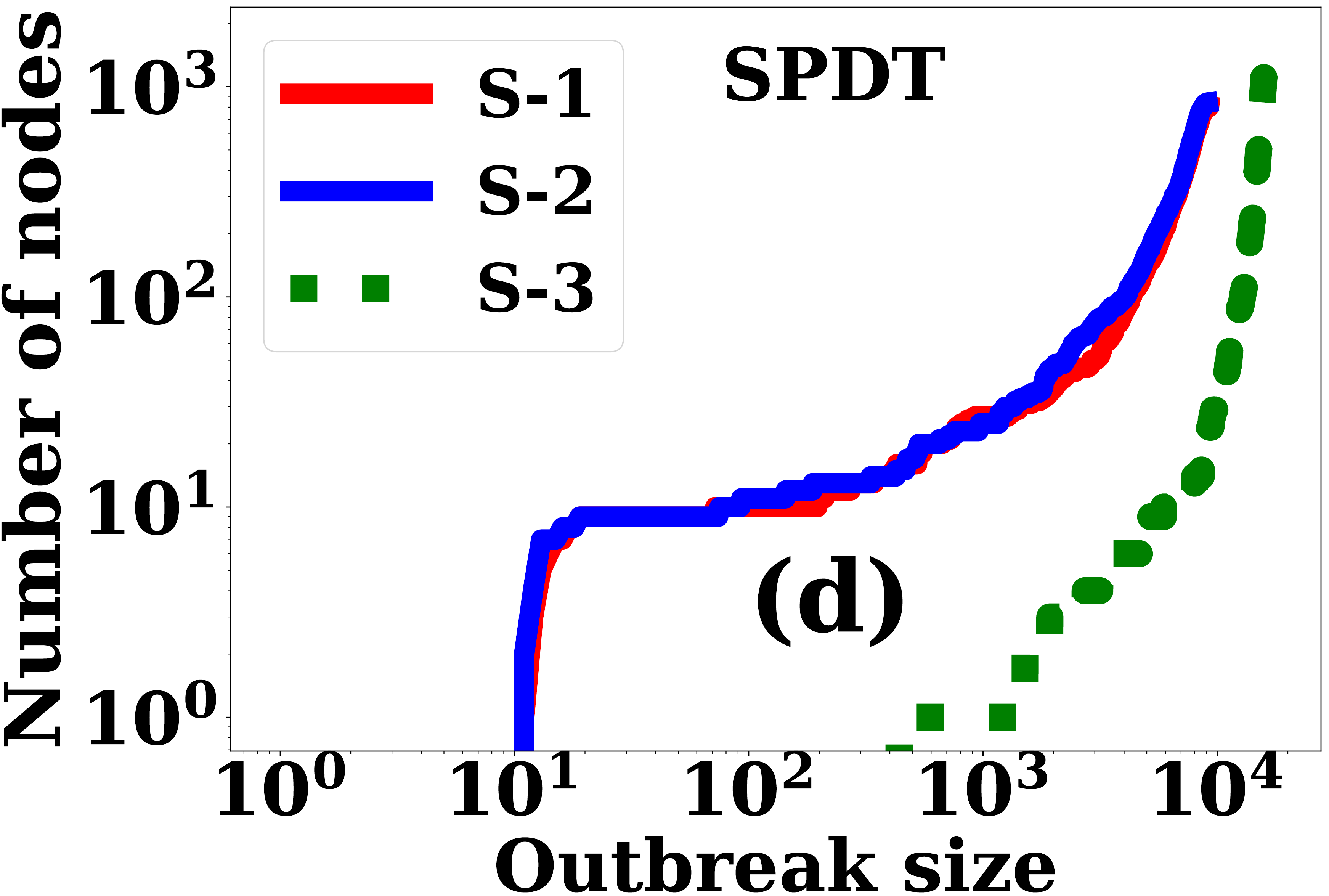}}
      \vspace{-0.5 em}
     \caption{Low connectivity nodes caused outbreaks: (a,b) number of nodes up to a specific disease prevalence and (c,d) number of nodes up to a outbreak size in both SPDT and SPST networks} 
     \label{fig:exp2}
\vspace{-1.9 em}
\end{figure}

The previous experiments show that the indirect links have strong impact on the  spreading behaviors of diseases. Now, we  study how these indirect links play vital roles at the individual level. To understand this, we investigate how the low direct connectivity nodes, that directly contact only one or two nodes during their infection periods through direct links, become important in SPDT network due to having the indirect links. We identify a low direct connection set of 10K nodes who have 1 or 2 neighbors in the SPST network over the first 5 days of our generated synthetic traces. Then, we run simulations by iterating through the nodes in this set to select each node as a seed node at $T=0$ on both SPST and SPDT networks. The seed nodes are able to infect others for 5 days before recovering from the infectious state. We keep the same mean and median of previous experiments respectively for b and g. We also run simulations for 3 scenarios (changing b, g and $\sigma$) to understand  how these nodes play a more significant role under certain conditions. 

The results are presented in Figure~\ref{fig:exp2} for the nodes that cause outbreak sizes greater than 10. In the first scenario {\bf S-1}, we set the parameters: mean of b is set to $0.01 min^{-1}$, median of $g$ to $1h^{-1}$ and $\sigma=0.69$. With this configuration, we find only 206 nodes can trigger disease in the SPST network. Comparatively, 803 nodes become capable to trigger disease in the SPDT network and their outbreak size is twice the outbreak sizes in SPST. If we change the value of g to $0.5 h^{-1}$ in scenario {\bf S-2}, the SPDT network allows 840 nodes to trigger disease while only one more node in the SPST network trigger disease. This is because the SPDT links are more pronounced for lower g while SPST links remain the same. If we make the scenario more favorable for spreading disease changing b to $0.005 min^{-1}$ and $\sigma=0.80$ (scenario {\bf S-3}), both networks offer more nodes to trigger disease where 307 nodes trigger disease in SPST network corresponding to 1649 nodes in the SPDT network. Under extreme conditions, the indirect links become significant, influencing the disease spreading strongly.
\begin{figure}
\centering
\vspace{-1.5 em}
	\subfloat{\includegraphics[width=0.35\textwidth, height=3.0 cm]{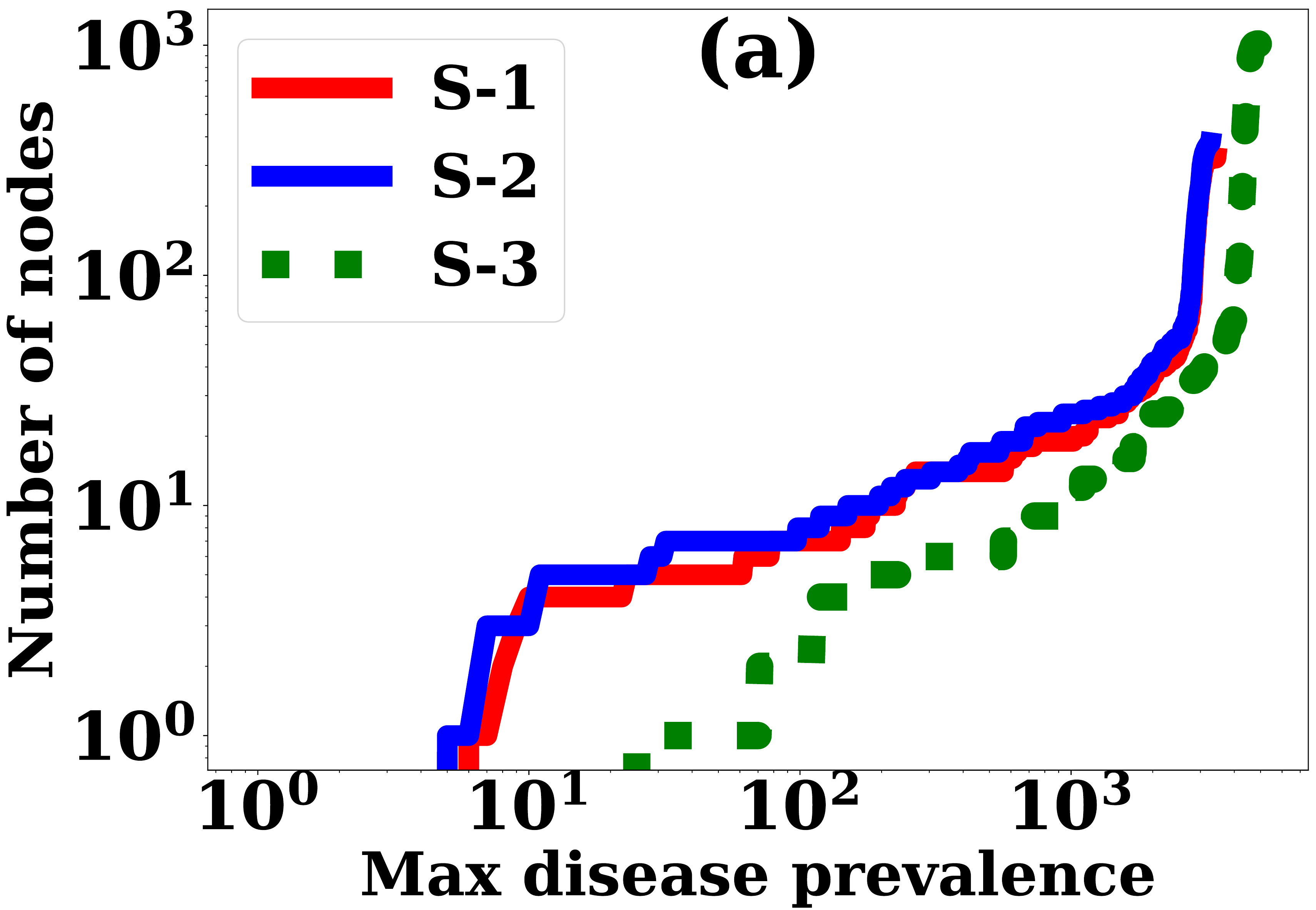}}~
    \subfloat{\includegraphics[width=0.35\textwidth, height=3.0 cm]{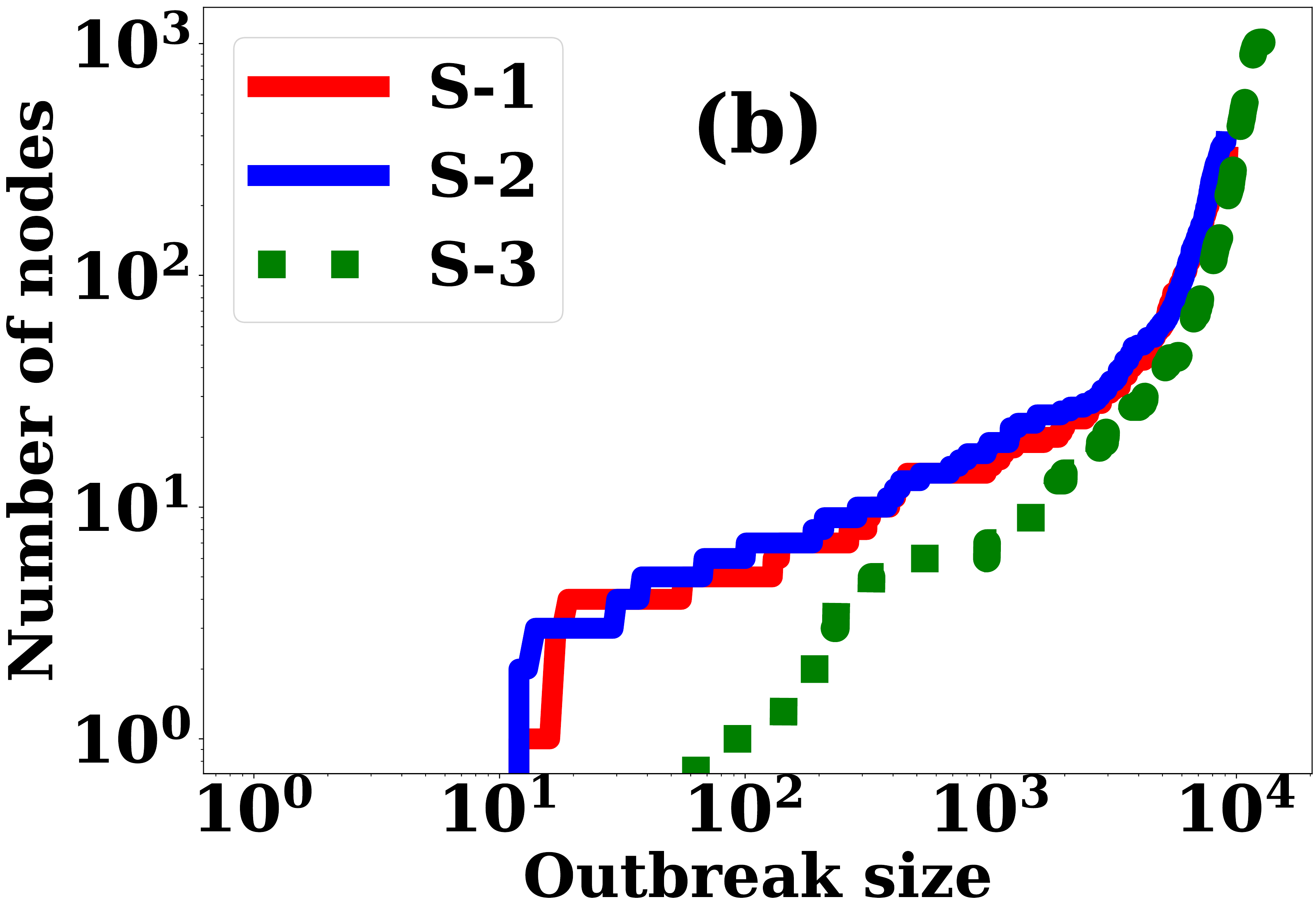}}
    \vspace{-0.5 em}
     \caption{Hidden spreaders (nodes with only indirect links during their infection periods) caused outbreaks: (a) number of nodes up to a disease prevalence and (b) number of nodes up to a specific outbreak size} 
     \label{fig:exp3}
     \vspace{-1.7 em}
\end{figure}

In the third experiment, we investigate whether single seed nodes that are hidden spreaders can trigger significant outbreaks. Using the hidden spreader set from the first 5 days from our first experiment, we study emergence of disease through all nodes selecting each as seed node at time $T=0$ of the simulations. The simulations are repeated 10 times for each node. We also explore whether the opportunities of emerging diseases are intensified for the favourable scenario of low b, g and high $\sigma$. The results are presented in Figure~\ref{fig:exp3}. For the first scenario, we find that 324 nodes cause outbreak sizes greater than 10 with maximum 7656 infections over the 32 day period. The medium spreading scenario allows 53 (16.3\%) more nodes to trigger diseases. If the diseases are more infectious having $\sigma=0.80$ in scenario S-3, 1014  nodes among the 11K nodes (which have only indirect links during the first 5 days of selected network traces) are capable to trigger disease. In this favorable spread scenario, the outbreak sizes as well as maximum prevalence of disease increase with earlier prevalence peaks.

%% file: discussion.tex
\section{Discussion and Future Work}
This paper has studied how the SPDT model enhances the opportunities of disease emergence due to  indirect links. These outcomes can guide more robust policy for controlling the infectious diseases. Our simulations on hidden spreaders have shown that indirect links are equally important as the direct links in driving the spread process. The indirect links increase the connectivity of the network and make it favorable for disease spreading. Running disease simulations from the low connectivity nodes has shown that 5 times more nodes can contract the disease if indirect links are considered. Interestingly, we noticed that nodes having only indirect links, hidden spreaders, can cause outbreaks with as many as 10K infections. However, these nodes in the SPST model can not infect others  as they do not have direct links. These outcomes require reconsideration of the disease spreading models of many infectious diseases.

Our work has some limitations. The infection risk is calculated assuming that the infection particles are distributed homogeneously in the space, which may not be true in reality, resulting in scaling of the infection rate and outbreak size. We only cover  particle removal by the air exchange rate and infectivity decay rate of infectious particles while particles be removed by other factors as well. The consideration of homogeneous susceptibility to a disease for individuals is not fully realistic as well. Secondly, we calculate the infection risk at the end of each simulation day for all links a node receives from the infected nodes. The infectiousness of the inhaled particle may vary over the course of the day in reality. These factors can over estimate infection rates and spreading dynamics. 

This work also opens some future research directions. It would be interesting to know why some nodes can initiate diseases. In our simulations, we observed that indirect links increase the connectivity among the nodes. This arises the question that the indirect links can lead nodes to become super-spreaders. We also observed favorable weather conditions caused larger outbreaks as the indirect links become  stronger. Another direction can be to investigate how the SPDT model with its capability to capture human mobility as well as environmental conditions can aid to design optimal diffusion control strategies. 